\documentclass{osa-article}
\usepackage{setspace}
\usepackage{color,soul}
\journal{oe}
\begin{document}

\title{Dual optical frequency combs with ultra-low relative phase jitters from 550 nm to 1020 nm for precision spectroscopy}

\author{Eunmi Chae,\authormark{1,2} Kota Nakashima,\authormark{3} and Kosuke Yoshioka\authormark{1,*}}

\address{
\authormark{1}Photon Science Center, School of Engineering, The University of Tokyo, Bunkyo-ku, Tokyo, Japan\\
\authormark{2}Department of Physics, Korea University, Seongbuk-gu, Seoul, Republic of Korea\\
\authormark{3}Department of Applied Physics,School of  Engineering, The University of Tokyo, Bunkyo-ku, Tokyo, Japan}

\email{\authormark{*}yoshioka@fs.t.u-tokyo.ac.jp} %% email address is required



\begin{abstract}
Here, ultra-low relative phase jitters over a wide optical spectrum were achieved for dual Ti:Sapphire optical frequency combs.
The two optical frequency combs were independently phase-locked to a Sr optical lattice clock laser delivered through a commercial optical fiber network. 
We confirmed that the relative phase jitters between the two combs integrated from 8.3 mHz to 200 kHz were below 1 rad, corresponding to a relative linewidth of below 8.3 mHz, over the entire wavelength of the optical frequency combs ranging from 550 nm to 1020 nm.
Our work paves the way for ultrahigh-precision dual-comb spectroscopy covering a wide optical spectral range with a simple setup, and provides an absolute optical frequency reference with great stability over a wide range of wavelengths.
\end{abstract}

%%%%%%%%%%%%%%%%%%%%%%%%%%  body  %%%%%%%%%%%%%%%%%%%%%%%%%%
\section{Introduction}
Ultra-precision spectroscopy plays a key role in modern science and technology, such as tests of fundamental physics, establishment of an optical frequency standard, and precision timing control for GPS and satellites \cite{Altmann2018,Biesheuvel2016,Webb2011,Huntemann2014,Godun2014,Chou2010,Lisdat2016,Derevianko2016,Roberts2017}.
A laser with a narrow linewidth and low phase noise is essential for ultra-precision spectroscopy.
Preparing an accurate optical reference over a broad wavelength range is necessary to meet the wavelength requirements of deep ultraviolet(UV) to far infrared(IR) applications.
The most advanced technique to reduce the phase noise of a laser is to lock the laser to a high-finesse cavity. 
The state-of-the-art laser has a sub-10 mHz linewidth when locked to a silicon Fabry-Pérot cavity at 124 K \cite{Matei2017}.
However, high-finesse cavities are available only at a few selected wavelengths.
Moreover, the setup for such an ultrastable laser requires a considerable amount of time and resources, including a temperature stabilized high-Q cavity in a shielded environment. 
Therefore, it would be ideal to distribute the phase coherence of an ultrastable optical reference over various locations and optical frequencies.

Recent progress in optical fiber networks has enabled the distribution of the stability of an ultrastable laser to remote locations with unprecedented precision.
In addition, by stabilizing an optical frequency comb (OFC) to the received reference laser, all longitudinal modes of the OFC, over several hundreds of THz, can acquire the same precision as the reference laser.
In previous research, we have shown that the stability of the optical reference can be reproduced with no degradation by the Ti:Sapphire OFC under challenging circumstances 1) in the presence of a 2.9 km-long fiber network and 2) without the local phase-purification cavity \cite{Chae2019}. 
A longitudinal mode of the OFC with optical frequency nearing that of the optical reference was used to examine the phase noise and stability of the OFC in the previous experiment.
However, the performance of other longitudinal modes, especially those with wavelengths far from that of the optical reference, remains unclear. 
Every optical frequency in a stabilized OFC is described as $f_n = f_\mathrm{CEO} + n \times f_\mathrm{rep}$ ($f_\mathrm{CEO}$, $f_\mathrm{rep}$, and $n$ are the carrier-envelope offset(CEO) frequency, the repetition frequency, and the mode number, respectively). 
In principle, the phase noises of all longitudinal modes of an OFC are tightly related to each other by this equation.
In practice, however, fluctuations of various physical origins can degrade the phase coherence of the modes that are away from the reference optical frequency.  
To validate the application of the OFC in ultra-precision spectroscopy over a wide wavelength range, the phase noise has to be characterized at various wavelengths.
In a previous research on Ti:Sapphire OFC, a sub-Hz residual linewidth was confirmed at the wavelength of 250 nm away from the optical reference \cite{Bartels2004}.
People have also demonstrated a fiber OFC with sub-Hz linewidths over the wavelength range of 1200 nm to 1720 nm \cite{Swann2006}.

Here, we studied the residual phase noise of the two OFCs by subjecting them to optical heterodyne detection, within the 550 -- 1020 nm wavelength range.
Two identical Ti:Sapphire OFCs were prepared and stabilized independently to a remotely located Sr optical lattice clock laser through an optical fiber network, as introduced in a previous study \cite{Chae2019}. 
Heterodyne beatnotes were detected at various wavelengths after overlapping the two pulse trains precisely in the time domain. 
The phase noise on the detected heterodyne beatnotes contains only the residual phase noise of the OFCs because the common noise of the optical reference and the fiber network cancel out. 
The residual linewidth of the OFCs over a wide range of wavelengths was analyzed based on the detected phase noise.

\section{Experimental Setup}
\begin{figure}[h]
    \centering
    \includegraphics[width=0.9\linewidth]{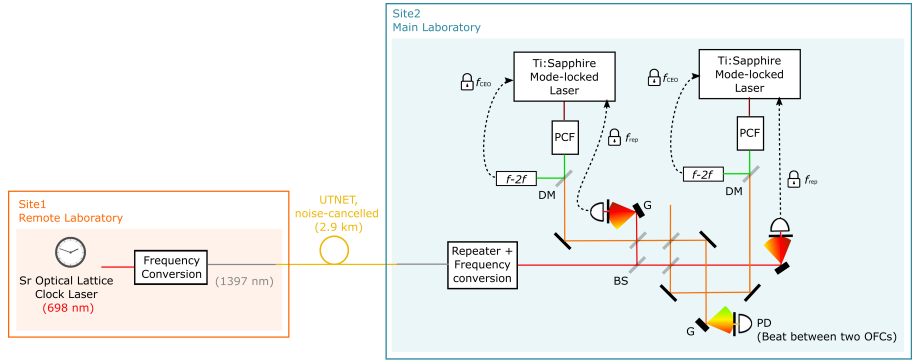}
    \caption{Overall experimental setup. The optical standard at the remote lab is delivered to the main lab through a telecom-wavelength fiber network (UT-NET) with fiber noise cancellation. Two identical OFCs, with self-referenced $f_{\mathrm{CEO}}$s, are phase-locked to the optical standard. The beat signal of the two OFCs is measured at various optical wavelengths to evaluate the residual phase noise of the OFCs. DM: dichroic mirror, BS: beam splitter, G: grating, PD: photo detector, PCF: photonic crystal fiber. }
    \label{fig:setup}
\end{figure}

The overall experimental setup consists of two Ti:Sapphire OFCs that are phase-locked to a remotely placed ultrastable laser for Sr optical lattice clock (optical standard) as described in ref. \cite{Chae2019}. 
To measure the heterodyne beat signal of the two OFCs, the timing of optical pulses from the two OFCs are precisely synchronized.
The residual linewidth of the OFC with respect to the optical standard is evaluated by analyzing the phase noise of the heterodyne beat signal. 

\subsection{Stabilization of the OFCs}
The setup of this experiment comprises two areas (Fig.\ref{fig:setup}).
We constructed two identical home-made Ti:Sapphire OFCs with output powers and repetition frequencies ($f_\textrm{rep}$) of approximately 500 mW and 117 MHz, respectively.
After broadening the wavelength range using photonic crystal fibers, the $f_{\mathrm{CEO}}$s of the OFCs were stabilized by a self-referencing technique with an $f$--$2f$ interferometer.
The optical standard, an ultranarrow-linewidth laser at 698 nm for a Sr optical lattice clock\cite{Takano2016}, was located at a laboratory (remote lab) about 1 km away from the location of the two OFCs (main lab). 
The optical standard was delivered to the main lab through a 2.9 km telecom-wavelength optical fiber network (UT-NET, University of Tokyo), phase noise of which was actively canceled\cite{Chae2019}. 
One of the frequency modes of the OFCs was phase-locked to the delivered optical standard with a frequency offset ($f_\mathrm{off}$). 
All RF signals and measurement apparatus were referenced by a GPS-disciplined Rb standard.
A detailed description of the stabilization and its performance is given in ref. \cite{Chae2019}.

In the previous experiment, the out-of-loop measurement at the wavelength near the optical standard detected the phase noise of 0.54 rad integrated from 8.3 mHz to 1.95 MHz with respect to the optical standard.   
However, this phase noise included high-frequency phase noise from the repeater laser of the optical standard.
The high-frequency phase noise cannot be transfered to the OFC due to the intrinsic narrow linewidth of the OFC. 
To correctly evaluate the phase noise of the OFC itself, we constructed two identical OFCs and evaluated the beat signal at various optical wavelengths.

\subsection{Heterodyne beatnote measurement of the two OFCs}
The optical paths of the two OFCs were carefully overlapped using a beam splitter so that the two light waves would interfere with each other.  
The full optical spectrum of the OFCs, 550 nm to 1100 nm, was used in the experiment.  
To reduce the phase noise due to different optical paths, we minimized the lengths of the out-of-loop path of the two OFCs.
Once the paths were combined, a grating was employed to separate the wavelengths of the OFCs.
The phase noise at selected wavelength ranges was measured by placing a slit between the grating and a photo detector, as shown in Fig. \ref{fig:setup}.
By rotating the grating, we changed the wavelength range to $549$ nm, $650$ nm, $750$ nm, $850$ nm, $948$ nm, and $1020$ nm.
The post-slit wavelength was measured using a spectrometer (Thorlabs, CCS200/M).

The pulse width of the spectrally resolved OFCs was approximately 100 fs.
For the two OFCs to interfere, a prerequisite is to synchronize the arrival times of the two OFC pulses within the pulse width.
In other words, the two OFCs should have the same $f_\mathrm{rep}$, and their repetition phases should match.
The frequency of the heterodyne beat ($f_\mathrm{beat}$) then appears at $f_\mathrm{CEO,1} - f_\mathrm{CEO,2}$.
To synchronize the timing of the pulses, we first chose the $f_\mathrm{CEO}$s and $f_\mathrm{off}$s, and match the $f_\mathrm{rep}$s of the two OFCs by properly chose the longitudinal mode locked to the optical reference.
We set $f_\mathrm{CEO,1} = 10.3$ MHz, $f_\mathrm{off,1} = -11.1$ MHz, $f_\mathrm{CEO,2} = -11.1$ MHz, and $f_\mathrm{off,2} = 10.3$ MHz, which resulted in $f_\mathrm{rep}=117$ MHz and $f_\mathrm{beat}=21.4$ MHz.

Matching the repetition phases is more challenging because the path-length of the two OFCs should be matched within a 30 $\mu$m - spatial spread of the pulses.  
To achieve this, we intentionally changed the $f_\mathrm{rep}$ of one OFC by a small amount ($< 10$ mHz), thereby changing the relative position between the two femtosecond pulses as a function of time. 
A heterodyne beat at 21.4 MHz appears when the two pulses overlap. 
At this moment, we immediately set $f_\mathrm{rep}$ of the OFC back to its original value.

\begin{figure}[h]
    \centering
    \includegraphics[width=0.7\linewidth]{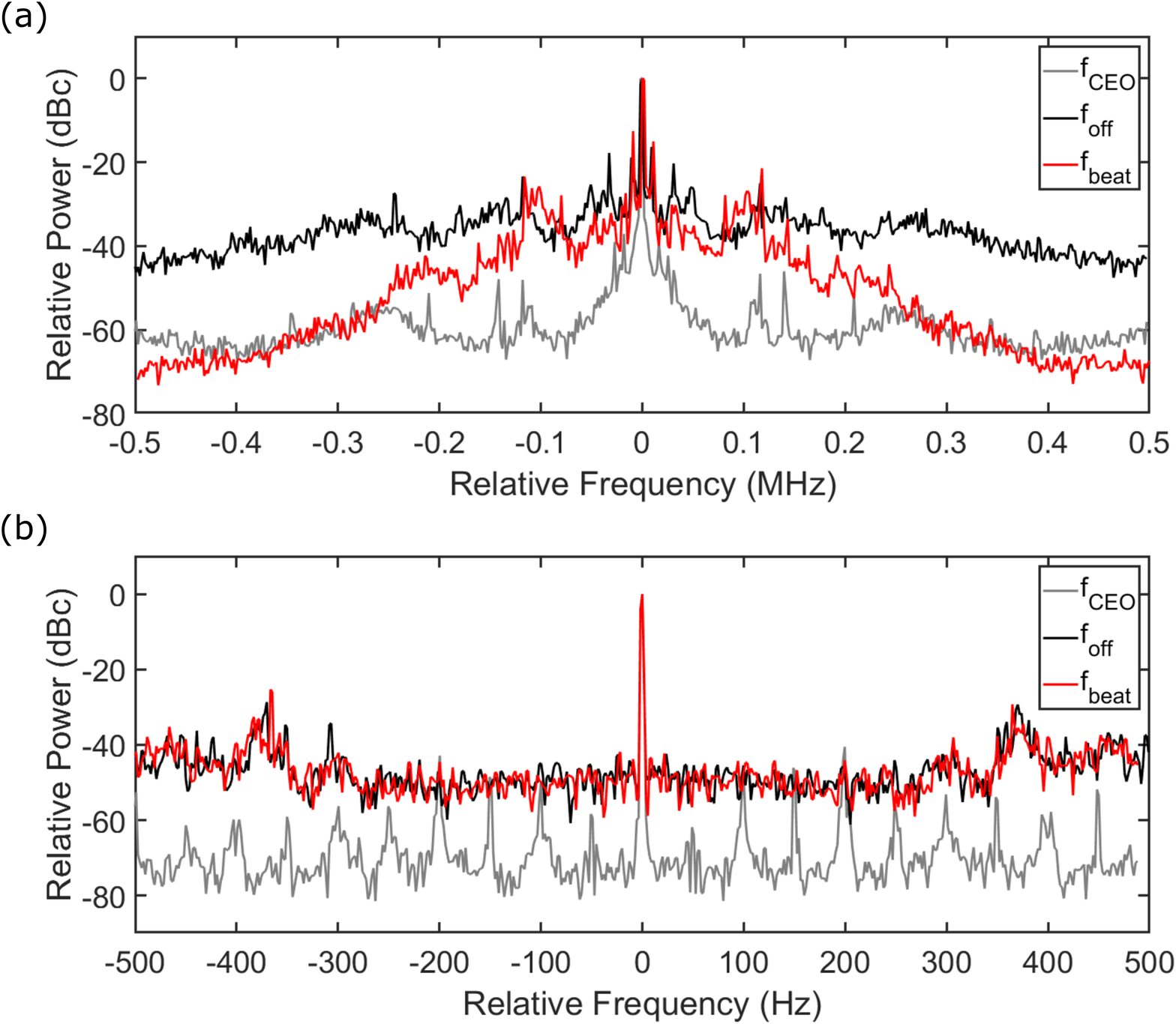}
    \caption{Heterodyne beat signal between the two OFCs at 650 nm. For reference, the in-loop spectra of $f_\mathrm{CEO}$ and $f_\mathrm{off}$ are also shown. (a) Resolution bandwidth of 300 Hz, (b) resolution bandwidth of 1 Hz.}
    \label{fig:combcombspectrum}
\end{figure}

\section{Results and Discussions}
The heterodyne beat signal at 650 nm, measured by a spectrum analyzer (Anritsu, MS2720T), is shown in Fig. \ref{fig:combcombspectrum} (red lines).
The signal was amplified by 50 dB using a home-made amplifier. 
The center frequency of the beat was 21.4 MHz as expected.
We observed a central coherent peak, the width of which is limited by the instrumental resolution. 
The phase noise at low frequencies matched very well the in-loop phase noise of $f_\mathrm{off}$.
However, at high frequencies, the heterodyne beat between the two OFCs had a significantly smaller phase noise. 
This indicates that the high frequency noise of $f_\mathrm{off}$ is from the repeater laser of the optical standard and is rejected during the active stabilization process of the OFCs.

The phase noise of the heterodyne beat was also measured after subtracting the carrier frequency to further examine the low frequency components of the phase noise.  
We compared the phase of the heterodyne beat to a reference RF signal generator using a home-made digital phase-frequency discriminator (DPFD). 
The output of the DPFD was measured using an oscilloscope (Pico Technology, PicoScope5244B) for 120 s with a sampling rate of 3. 9 MHz.
The two OFCs were phase-locked during the 120 s measurement without any phase jumps.  
However, the output power fluctuation of the OFCs caused the phase signal of the resulting beat to be discontinuous, at wavelength of 550 nm, 850 nm, 950 nm, and 1020 nm.
The discontinuities were especially significant for 550 nm and 1020 nm, corresponding to the edges of the OFC optical spectra.
This occurred when the input signal voltage was insufficient relative to the threshold voltage of the DPFD.
In order to concentrate only on the phase noise, we selected time windows where there were no discontinuous phase signals for each wavelength.

The power spectral density (PSD) of the phase noise obtained from the raw data had a spectral range of 8.3 mHz to 1.95 MHz.
We confirmed that there were negligible amount of phase noise at frequencies higher than 20 kHz due to the intrinsic narrow linewidth of our OFCs. 
Therefore, the PSD components upto 20 kHz were used for the following analysis. 

The measured phase noise had a linear drift on the order of 0.003 rad/s, with magnitude and sign varing over time.
The drift is due to the different optical paths between the OFCs.
The drifts, as they are not essential for our discussion, are excluded in the following analysis.

\begin{figure}[h]
    \centering
    \includegraphics[width=0.7\linewidth]{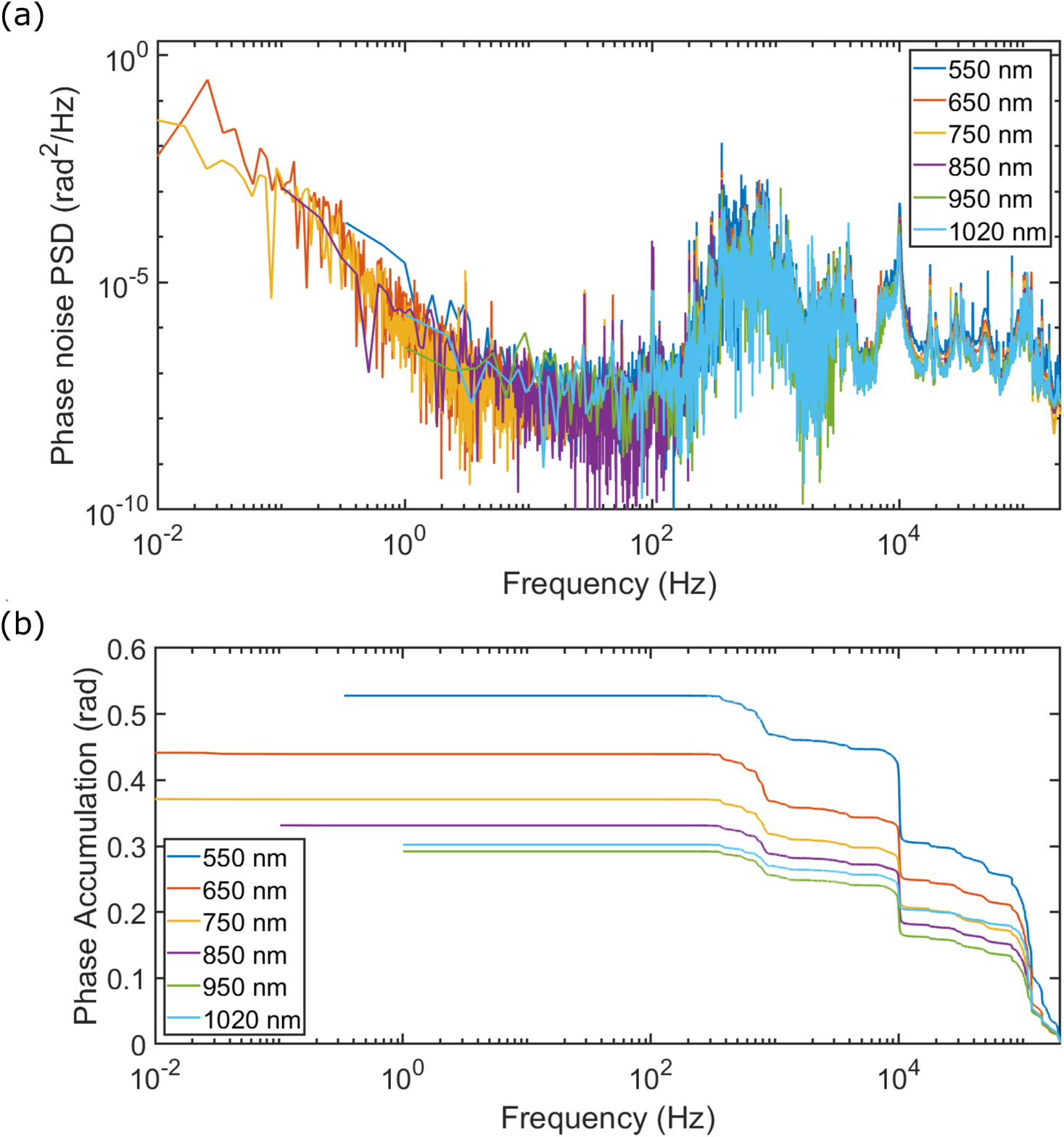}
    \caption{(a) PSD and (b) integrated phase jitters of the phase noise of the heterodyne beat between the two OFCs at each wavelength.}
    \label{fig:PhasePSD}
\end{figure}

The PSD of the phase noise of the beat between the two OFCs is shown in Fig. \ref{fig:PhasePSD} (a).
All PSDs display identical shapes over the entire wavelength range of over an octave of the OFCs, indicating that no additional phase noise was added to the OFCs.
Below 1 Hz, the PSDs show the typical $1/f^2$ dependence of white frequency modulation noise.
Fig. \ref{fig:PhasePSD} (b) shows the accumulated phase noise of each OFC calculated from the PSDs. 
The accumulated phase is smaller than 1 rad over the whole wavelength range. 
The residual linewidth of the OFC is smaller than the resolution of our measurement, namely 8.3 mHz over the whole wavelength range.
The relative linewidth demonstrated here is two orders of magnitudes narrower than the previous works \cite{Bartels2004, Swann2006}.

\begin{figure}[h]
    \centering
    \includegraphics[width=0.7\linewidth]{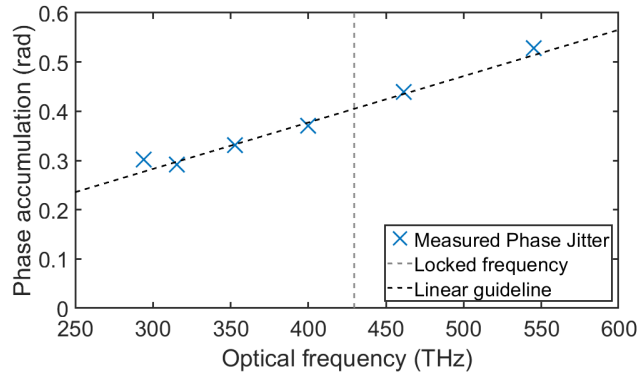}
    \caption{Accumulated phase jitter at 1 Hz for various wavelengths. The black dashed line indicates the linear guideline. The y intercept of this line is zero. The frequency of the optical standard is represented by the gray vertical dashed line. }
    \label{fig:AccuPhase}
\end{figure}

When the $f_\textrm{CEO}$ and $f_\textrm{rep}$ of an OFC are stabilized, the phase accumulation should increase linearly as a function of the optical frequency of the OFC \cite{Swann2006}.
As shown in Fig. \ref{fig:AccuPhase}, the accumulated phase jitters at 1 Hz follow a linear function over most of the optical spectrum of the OFC.
The accumulated phase jitter at 298 THz ($= 1020$ nm) is more pronounced than expected because of the optical power fluctuation.
The timing jitter of the OFC pulses is estimated to be approximated 150 as from the accumulated phase jitter. 

The two tightly phase-locked OFCs with absolute frequency accuracy are excellent tools for the precision dual-comb spectroscopy \cite{Coddington2016}.
The optical bandwidth and the frequency accuracy are the two features that describe the performance of the dual comb spectroscopy. 
The optical bandwidth of the dual-comb spectroscopy is determined by the ratio between the relative linewidth of the two OFCs and $f_\textrm{rep}$.
Therefore, it is important that the OFCs maintain the narrow relative linewidth over a wide optical wavelength range to achieve a large optical bandwidth.
The frequency resolution and accuracy are assured when the two OFCs have absolute frequency modes precisely described by $f_n = f_\mathrm{CEO} + n \times f_\mathrm{rep}$ with narrow linewidth over the whole spectral range. 
Our result with < 1 rad phase jitters over 120 s for an octave-spanning wavelength range would allow dual comb spectroscopy with an unprecedented precision.

\begin{figure}[h]
    \centering
    \includegraphics[width=\linewidth]{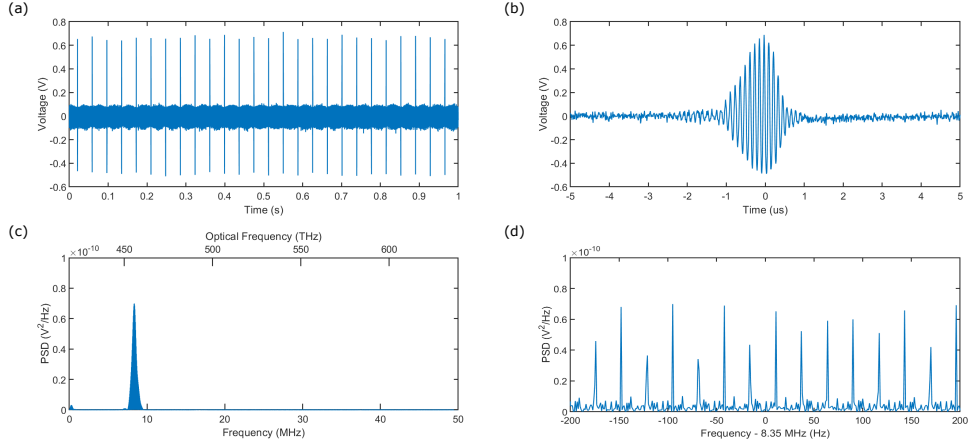}
    \caption{Demonstration of the dual-comb spectroscopy using our OFCs. Owing to the long coherence time of the two OFCs, RF spectra with a 1 Hz linewidth could be obtained. (a)(b): The raw interference data. (c)(d) The RF spectrum obtained by Fourier transform. Every RF mode is separated by 26.45 Hz and the linedwidth of 1 Hz is set by the measurement time.}
    \label{fig:Dual}
\end{figure}

Fig. \ref{fig:Dual} presents an example of a dual-comb spectroscopy signal using our OFCs.
We selected wavelengths in the 655 to 665 nm range (optical frequency with of 7.2 THz, approximately 60,000 frequency modes) using a band-pass filter for this test.
We set the $\Delta f_\textrm{rep} = 26.45$ Hz in this measurement.
Fig. \ref{fig:Dual} (a) is the raw data taken by a photo detector with 150 MHz bandwidth and a 16-bit digitizer with a 100 MHz sampling frequency.
The interference signals of the two OFCs appear every $1/26.45 \sim  40$ ms.
By zooming into Fig. \ref{fig:Dual} (b), the beat inside the pulse can be observed. 
The RF spectrum of the signal is obtained by performing a Fourier transform of the data after filtering out the noise during periods of no interference (Fig. \ref{fig:Dual} (c)). 
The width of the RF spectrum is approximately 1 MHz, which is comparable to the expected value of $26.45 \times 60,000 \sim 1.6$ MHz.
Fig. \ref{fig:Dual} (d) shows the zoomed in spectrum where one can clearly resolve each mode of the RF comb separated by 26.45 Hz with the linewidth of 1 Hz limited by our measurement time of 1 s. 
This confirms that our OFCs maintain their coherence over 1 s.

\section{Conclusion}
We constructed two Ti:Sapphire OFCs that were tightly phase-locked to each other over a wide wavelength range, 550 nm to 1020 nm. 
The two OFCs were independently phase-locked to an optical lattice clock at 698 nm via an optical fiber network. 
The phase noise at different wavelengths was measured by taking heterodyne beatnotes of the two OFCs.
The resulting phase jitters displayed a linear increase as a function of the optical frequencies, which is in agreement with the theoretical prediction. 
Phase jitters of less than 1 rad integrated from 8.3 mHz to 20 kHz indicate that the two OFCs are tightly phase-locked to each other over the entire wavelength range of 550 nm to 1020 nm. 
The phase jitter measured here corresponds to a relative linewidth of < 8.3 mHz and a timing jitter of 150 as. 
Our result shows that these OFCs transfer the coherence of a remote optical clock laser with minimal phase noise to arbitraty wavelengths in the range stated above. 
Therefore, the referenced OFCs with the current setup can be utilized for state-of-the-art precision spectroscopy over a wide wavelength range. 
Also, the ultrastable OFC demonstrated here is an excellent tool for dual-comb spectroscopy covering a wide optical bandwidth with a Hz-level resolution and a pump-probe measurement with sub-fs timing resolution.
It also provides an absolute optical frequency reference with great stability over a wide range of wavelengths.

\begin{backmatter}
\bmsection{Funding}
This work is supported by Kakenhi (JP17H06205, JP20K05357), MEXT Quantum Leap Flagship Program (MEXT QLEAP) Grant No. JPMXS0118067246, PRESTO, JST (JPMJPR190B). E. C. is supported by NRF of Korea (2020R1A4A1018015, 2021R1C1C1009450) and a Korea University Grant (K2009521).

\bmsection{Acknowledgments}
The authors thank H. Katori for providing access to the clock laser and other equipment in his laboratory. 
The authors appreciate the technical support provided by T. Takano, N. Ohmae, T. Akatsuka and I. Ushijima. 
The authors thank M. Kuwata-Gonokami for presenting the
conceptual idea in Advanced Photon Science Alliance (APSA) project that led to the present work, and providing basic equipment at the OFC site.  

\bmsection{Disclosures}
The authors declare no conflicts of interest.

\bmsection{Data availability} Data underlying the results presented in this paper are not publicly available at this time but may be obtained from the authors upon reasonable request.

\end{backmatter}

%%%%%%%%%%%%%%%%%%%%%%% References %%%%%%%%%%%%%%%%%%%%%%%%%

%%%%%%%%%% If using BibTeX:
\bibliography{References-DualComb}

\end{document}